# Econophysics: A Brief Review of Historical Development, Present Status and Future Trends.

B.G.Sharma
Department of Physics and Computer Science,
Govt. Science College Raipur. (India)
Sharma_balgopal@rediffmail.com

Sadhana Agrawal
Department of Physics
NIT Raipur. (India)
agrawaldrs@rediffmail.com

Malti Sharma
WQ-1, Govt. Science College
Raipur. (India)
Sharma_balgopal@rediffmail.com

D.P.Bisen
SOS in Physics,
Pt. Ravishankar Shukla University
Raipur. (India)
dpbisen@rediffmail.com

Ravi Sharma
Devendra Nagar Girls College
Raipur. (India)
rvsharma65@gmail.com

**Abstract**: The conventional economic approaches explore very little about the dynamics of the economic systems. Since such systems consist of a large number of agents interacting nonlinearly they exhibit the properties of a complex system. Therefore the tools of statistical physics and nonlinear dynamics has been proved to be very useful the underlying dynamics of the system. In this paper we introduce the concept of the multidisciplinary field of econophysics, a neologism that denotes the activities of Physicists who are working on economic problems to test a variety of new conceptual approaches deriving from the physical science and review the recent developments in the discipline and possible future trends.

**Broad Area**: Physics

**Sub Area**: Econophysics

## 1. Introduction:

How is the stock market like the cosmos or like the nucleus of an atom? To a conservative physicist, or to an economist, the question sounds like a joke. It is no laughing matter, however, for Econophysicists seeking to plant their flag in the field of economics. In the past few years, these trespassers have borrowed ideas from quantum mechanics, string theory, and other accomplishments of physics in an attempt to explore the divine undiscovered laws of finance. They are already tallying what they say are important gains. The tools of physics provide an ideal background for approaching problems in economics [1]. Physics training, gives a person powerful mathematical tools, computer savvy, a facility in manipulating large sets of data, and an intuition for modeling and simplification. Such skills have brought a new order into economics.

## 2. What lies in Econophysics?

**Econophysics** is an interdisciplinary research field, applying theories and methods originally developed by physicists in order to solve problems in economics, usually those including uncertainty or stochastic processes and nonlinear dynamics. Its application to the study of financial markets has also been termed statistical finance referring to its roots in statistical physics. Physics has played an important role in the development of economic theory through the 19th century, and some of the founders of neoclassical economic theory, were originally trained as physicists.

## 2.1 Why Econophysics?

The quantitative success of the economic sciences is disappointing when it is compared with that of physics. Its recurrent inability to predict and avert crises, including the current worldwide credit crunch is obvious? Why is this so? Of course, modeling the madness of people is more difficult than the motion of planets, as Newton once said. But the goal here is to describe the behavior of large populations, for which statistical regularities should emerge. The crucial difference between physical sciences and economics or financial mathematics is rather the relative role of concepts, equations and empirical data. Classical economics is built on very strong assumptions that quickly become axioms: the rationality of economic agents, the invisible hand and market efficiency, etc. Physicists, on the other hand, have learned to be suspicious of axioms and models. If empirical observation is incompatible with the model, the model must be trashed or amended, even if it is conceptually beautiful or mathematically convenient. So many accepted ideas have been proven wrong in the history of physics that physicists have grown to be critical and queasy about their own models. Unfortunately, such healthy scientific revolutions have not yet taken hold in economics, where ideas have solidified into dogmas. In reality, markets are not efficient, humans tend to be over-focused in the short-term and blind in the long-term, and errors get amplified through social pressure and herding, ultimately leading to collective irrationality, panic and crashes. Free markets are wild markets. It is foolish to believe that the market can impose its own self-discipline.

Reliance on models based on incorrect axioms has clear and large effects [2]. The Black-Scholes model assumes that price changes have a Gaussian distribution, i.e. the probability of extreme events is deemed negligible. Unwarranted use of this model to hedge the downfall risk on stock markets spiraled into the October 1987 crash. Ironically, it is the very use of the crash-free Black-Scholes model that destabilized the market! In the recent subprime crisis of 2008 also, the problem lay in part in the development of structured financial products that packaged sub-prime risk into seemingly respectable high-yield investments. The models used to price them were fundamentally flawed: they underestimated the probability of the multiple borrowers would default on their loans simultaneously. In other words, these models again neglected the very possibility of a global crisis, even as they contributed to triggering one. Surprisingly, there is no framework in classical economics to understand wild markets, even though their existence is so obvious to the layman. Physicists, on the other hand, has developed in physics, several models allowing one to understand how small perturbations can lead to wild effects. The theory of complexity, developed in the physics literature over the last thirty years, shows that although a system may have an optimum state (such as a state of lowest energy, for example), it is sometimes so hard to identify that the system in fact never settles there. This optimal solution is not only elusive, it is also hyper-fragile to small changes in the environment, and therefore often irrelevant to understanding what is going on. There are good reasons to believe that this complexity paradigm should apply to economic systems in general and financial markets in particular. Simple ideas of equilibrium and linearity do not work. We need to break away from classical economics and develop altogether new tools, as attempted in a still patchy and disorganized way by behavioral economists and econophysicists. But their fringe endeavour is not taken seriously by mainstream economics.

Thus there is a crucial need to change the mindset of those working in economics and financial engineering. They need to realize that an overly formal and dogmatic education in the economic sciences and financial mathematics is serious part of the problem. In sum the Economic curriculums need to include more natural science so that it can tackle the real world problems more accurately and efficiently.

3## 2.2 Historical Development:

Econophysics studies were started in the mid 1990s by several physicists working in the subfield of statistical mechanics [3-5]. They decided to tackle the complex problems posed by economics, especially by financial markets. Unsatisfied with the traditional explanations of economists, they applied tools and methods from physics - first to try to match financial data sets, and then to explain more general economic phenomena. With the availability of huge amounts of financial data, starting in the 1980s, it became apparent that traditional methods of analysis were insufficient. Standard economic methods dealt with homogeneous agents and equilibrium, while many of the interesting phenomena in financial markets fundamentally depended on heterogeneous agents and far-from-equilibrium situations.

The term "econophysics" was coined by H. Eugene Stanley in the mid 1990s, to describe the large number of papers written by physicists in the problems of stock and other markets, and first appeared in a conference on statistical physics in Calcutta in 1995 and its following publications. The inaugural meeting on Econophysics was organised 1998 in Budapest by János Kertész and Imre Kondor.

Though the term "Econophysics" has been entered the scientific language only about one and half decade ago, the connection and interplay between physics and economy are about 300 years old [6-8]. Literature is full of examples of famous physicist's involvement in economic or financial problems. Daniel Bernoulli introduced the idea of utility to describe people's preferences (1738). Pierre-Simon Laplace, in his *Essai philoso-phique sur les probabilites* pointed out that events that might seem random and unpredictable in economics can be quite predictable and can be shown to obey simple laws (1812). Adolphe Quetelet further amplified the Laplace's ideas by studying the existence of patterns in data sets ranging from economic to social problems. (1835). Irving Fisher, originally trained as physicists, and a student of Willard Gibbs played an important role in the development of neoclassical economic theory. The first formalism of random walk (a mathematical model of efficient markets) was not in a publication by Einstein, but in Doctoral thesis by Luis Bachelier. His work dealt the first formulation of the pricing of options in speculative markets, an activity that is extremely important in fancial markets (1900). In 1938, Ettore Majorana pre-sciently outlined both the opportunities and pitfalls in applying statistical physics method to socio economic systems. Jan Tinbergen, who studied physics with Paul Ehrenfest at Leiden University, won the first Nobel Prize in economics in 1969 for having developed and applied dynamic models for the analysis of economic processes. Ingrao and Israel showed that the works of Léon Walras and Vilfredo Pareto on equilibrium economics is, in fact, based on the physical concept of mechanical equilibrium. One of the most revolutionary development in the theory of speculative prices since Bachelier's initial work, is the Mandelbrot's hypothesis that price changes follow a Levy stable distribution rather than a Gaussian one. A widely accepted belief in financial theory is that time series of asset prices are unpredictable. Poincare (1854-1912) has pointed the possibility of unpredictability in a. nonlinear dynamical system, establishing the foundations of the chaotic behavior. The study of chaos turned out to be a major branch of theoretical physics. It was only a question of time, how fast these ideas will start to appear in economy. Ironically, Poincare, who did not appreciate Bachelier's results, made himself a large impact on real complex systems as one of the discoverers of chaotic behavior in dynamical systems. Nowadays studies of chaos, self-organized criticality, cellular automata and neural networks are seriously taken into account, as economical and financial tools.

The next major factor, changing the Gaussian world was a computer. First, it has changed the speed and the range of transactions drastically. The application of computer started involuntarily to serve as an amplifier of fluctuations. Second, the economies and markets started to watch each other more closely, since computer possibilities allowed for collecting exponentially more data. In this way, several nontrivial couplings started to appear in economical systems, leading to nonlinearities. Nonlinear behavior and overestimation of the Gaussian principle for fluctuations were responsible for the Black Monday Crash in 1987, and the crisis in August, and September 1998 and sub-prime crisis of 2008. That shock had however also a positive impact visualizing the- importance of the non-linear effects. Poincare has long ago pointed the possibility of unpredictability in a nonlinear dynamical system, establishing the foundations of the chaotic behavior. The study of chaos turned out to be a major branch of theoretical physics. It was only a question of time, how fast these ideas will start to appear in economy. Ironically, Poincare, who did not appreciate Bachelier's results, made himself a large impact on real complex systems as one of the discoverers of chaotic behavior in dynamical systems. Nowadays studies of chaos, self-organized criticality, cellular automata and neural networks are seriously taken into account, as economical and financial tools. One of the benefits of the computers was that economic systems started to save more and more data. Today markets collect incredible amount of data. This triggers the need for new methodologies, able to manage the data. In particular, the data started to be analyzed using methods, borrowed widely from physics, where seeking for regularities and for unconventional correlations is mandatory. In the last fifteen years, several educational and research institutions devoted to study of complexity launched the research programs in economy and financial engineering. These studies were devoted mostly to quantitative finance. To a large extent, it was triggered by vast amount of data accessible in this field

## 2.3 Present Status:

Recently, a growing number of physicists have attempted to analyze and model financial markets and, more generally, economic systems [6]. This unorthodox point of view was considered of marginal interest until recently. Indeed, prior to the 1990s, very few professional physicists did any research associated with social or economic systems. Since 1990, the physics research activity in this field has become less episodic and a research community has begun to emerge. The research activity of this group of physicists is complementary to the most traditional approaches of finance and mathematical finance. One characteristic difference is the emphasis that physicists put on the empirical analysis of economic data. Another is the background of theory and method in the field of statistical physics developed over the past three decades that physicists bring to the subject. The concepts of scaling, universality, disordered frustrated systems, and self-organized systems might be helpful in the analysis and modeling of financial and economic systems. Financial firms on Wall Street in U.S.A. put out welcome mats for physicists over a decade ago. People with physics Ph.D.s hold about half of the so-called quantitative analyst positions at such institutions, and they significantly outnumber economists. Wall Street physics has been mostly a proprietary pursuit of new spins on old methods for concocting abstract financial instruments, of which stock options are among the simplest examples. In the margins, a few physicist-financiers are working on so-called black box trading schemes.

Now, the embrace of physics and finance has been reached into academics. Physicists at universities are taking up finance, and nonacademic physicists in finance are pursuing basic research. Together they published about more than 100 economics papers last year in journals of physics and the number is increasing exponentially yoy. Currently, the almost regular meeting series on the topic include: Econophysics Colloquium, ESHIA/ WEHIA, ECONOPHYS-KOLKATA, APFA. Participants in the movement say that research in finance is growing faster than in any other area of physics. Within the dark recesses of proprietary financial research on and off Wall Street, an unreckoned but purportedly small number of stock analysts are building what they call black boxes. These computerized systems monitor current and past prices of a stock or asset, consult currency exchange rates or other factors that might serve as financial indicators, and spit out decisions from moment to moment about whether an investor should buy or sell. An ever-evolving formula instructs some black boxes as to which indicators to consult and how to factor them into the decisions. The boxes themselves may devise these formulas. Many boxes evaluate stocks using programming that mirrors how brain-cell networks operate. The computers effectively teach themselves as they go along how to forecast swings in price. However, the goal of black box research is narrow. Researchers have a strong disincentive to publish any innovations that would be useful for turning profits, since if everybody knew of them, they would cease to work.

This brief presentation of some of the current efforts in this emerging discipline can be summarized as follows:

### 2.3.1. Statistical characterization of the stochastic process of price changes of a financial asset:

Among the important areas of physics research dealing with financial and economic systems, one concerns the complete statistical characterization of the stochastic process of price changes of a financial asset. Several studies have been performed that focus on different aspects of the analyzed stochastic process, e.g., the shape of the distribution of price changes, the temporal memory, and the higher-order statistical properties . This is still an active area, and attempts are ongoing to develop the most satisfactory stochastic model describing all the features encountered in empirical analyses. One important accomplishment in this area is an almost complete consensus concerning the finiteness of the second moment of price changes. This has been a longstanding problem in finance, and its resolution has come about because of the renewed interest in the empirical study of financial systems.

### 2.3.2 development of a theoretical model:

A second area concerns the development of a theoretical model that is able to encompass all the essential features of real financial markets. Several models have been proposed , and some of the main properties of the stochastic dynamics of stock price are reproduced by these models as, for example, the leptokurtic 'fat-tailed' non-Gaussian shape of the distribution of price differences. Parallel attempts in the modeling of financial markets have been developed by economists.

### 2.3.3. Rational pricing of a derivative products:

Other areas that are undergoing intense investigations deal with the rational pricing of a derivative product when some of the canonical assumptions of the Black & Scholes model are relaxed and with aspects of portfolio selection and its dynamical optimization. A further area of research considers analogies and differences between price dynamics in a financial market and such physical processes as turbulence and ecological systems .

### 2.3.4. Time correlation of a financial series:

One common theme encountered in these research areas is the time correlation of a financial series. The detection of the presence of a higher-order correlation in price changes has motivated a reconsideration of some beliefs of what is termed 'technical analysis.

### 2.3.5. Income distribution of firms and their growth :

In addition to the studies that analyze and model financial systems, there are studies of the income distribution of firms and studies of the statistical properties of their growth rates. The statistical properties of the economic performances of complex organizations such as universities or entire countries have also been investigated .

### 3. Impact on mainstream economics and finance:

Papers on econophysics have been published primarily in journals devoted to physics and statistical mechanics, rather than in leading economics journals. Mainstream economists have generally been unimpressed by this work. Some Heterodox economists, including Mauro Gallegati, Steve Keen and Paul Ormerod, have shown more interest, but also criticized trends in econophysics.
In contrast, econophysics is having some impact on the more applied field of quantitative finance, whose scope and aims significantly differ from those of economic theory. Many econophysicists have introduced models for price fluctuations in financial markets or original points of view on established models.

### 4. Conclusion :

Knowledge of dynamical properties of economic systems is essential for fundamental and applied reasons. Such knowledge is crucial for the building and testing of a model of economic market. Dynamics enters the economics in two quite different and fundamental ways. The first, which has its counterpart in the natural sciences, is from the fact that the present depends upon the past. Such models typically are of the form

$$y_t = f(y_{t-1})$$

where we consider just a one period lag (a Markov process). The second way dynamics enters macroeconomics, *which has no counterpart in the natural sciences,* arises from the fact that economic agents in the present have expectations (or beliefs) about the future. Again taking a one-period analysis, and denoting the present expectation about the variable *y* one period from now by $E\,y_{t+1}$, then

$$y_, = g(Ey_{t+1})$$

Let us refer to the first lag as a *past lag* and the second *a future lag.* There is certainly no reason to suppose modeling past lags is the same as modeling future lags. Furthermore, a given model can incorporate both past lags and future lags. The natural sciences provide the mathematics for handling past lags but has nothing to say about how to handle future lags. It is the future lag which gained most attention in the 1970s, most especially with the rise in rational expectations. Once a future lag enters a model it becomes absolutely essential to model expectations, and at the present time there is no generally accepted way of doing this. This does not mean that we should not model expectations, rather it means that at the present time there are a variety of ways of modeling expectations, each with its strengths and weaknesses. This is an area for future research. In spite of a long effort, this goal has not yet been achieved. Statistical and Theoretical physicists can contribute a lot to the resolution of these scientific problems by sharing , with researchers in the other disciplines involved, the background in critical phenomena, disordered systems, scaling, and universality that has been developed over last 30 years. Despite the field's long history of association, the substantial contribution of physics to economics is still in an early stage, and we think it fanciful to predict what will ultimately be accomplished. Almost certainly, "physical" aspects of theories of social order will not simply recapitulate existing theories in physics, though already there appear to be overlaps. The development of economies can be contingent on accidents of history and at every turn hinges on complex aspects of human behavior.

Nonetheless, striking empirical regularities suggest that at least some social order is not historically contingent, and is perhaps predictable from first principles. The role of markets as mediators of communication and distributed computation, which underlie the collective processes of price formation and allocation of resources, and the emergence of the social institutions that support those functions, are quintessentially economic phenomena. Yet the notions of markets' communication or computational capacities, and the way differences in those capacities account for the stability and historical succession of markets, may naturally be part of the physical world with its human social dynamics. Markets and other economic institutions bring with them concepts of efficiency or optimality in satisfying human desires. While intuitively appealing, such ideas have proven hard to formalize even if some progress has been made. As with most new areas of physical inquiry, we expect that the ultimate goals of a physical economics will be declared with hindsight, from successes in identifying, measuring, modeling, and in some cases predicting empirical regularities. One argument that is sometimes raised at this point is that an empirical analysis performed on financial or economic data is not equivalent to the usual experimental investigation that takes place in physical sciences. In other words, it is impossible to perform large-scale experiments in economics and finance that could falsify any given theory.

We note that this limitation is not specific to economic and financial systems, but also affects such well developed areas of physics as astrophysics, atmospheric physics, and geophysics. Hence, in analogy to activity in these more established areas, we find that we are able to test and falsify any theories associated with the currently available sets of financial and economic data provided in the form of recorded files of financial and economic activity.

**Appendix**:

***Some Important Centers of Econophysics Research*:**

1. Boston University USA.
2. Santa Fe Institute USA
2. Saha Institute of Nuclear Physics, Kolkata, India.
3. Ecole Centrale Paris, France.
4. University of Maryland, UK.
5. University of Palermo, Italy.